\documentclass[reqno, 10pt]{amsart}

\usepackage{color}
\usepackage{mathrsfs}
\usepackage{amssymb}

\newtheorem{theorem}{Theorem}[section]
\newtheorem{corollary}[theorem]{Corollary}
\newtheorem{lemma}[theorem]{Lemma}
\newtheorem{proposition}[theorem]{Proposition}
\theoremstyle{definition}
\newtheorem{definition}[theorem]{Definition}
\theoremstyle{remark}
\newtheorem{remark}[theorem]{Remark}
\newtheorem{example}[theorem]{Example}
\numberwithin{equation}{section}

\newcommand{\refcite}[1]{\cite{#1}}

\begin{document}

\title{The stability of quantum Markov filters}
\author{Ramon van Handel}
\address{California Institute of Technology 266-33\\
Pasadena, CA 91125, USA}
\email{ramon@its.caltech.edu}
\thanks{This work is supported by the ARO under grant W911NF-06-1-0378.}

\begin{abstract}
When are quantum filters asymptotically independent of the initial state?
We show that this is the case for absolutely continuous initial states 
when the quantum stochastic model satisfies an observability condition.
When the initial system is finite dimensional, this condition can be 
verified explicitly in terms of a rank condition on the coefficients of 
the associated quantum stochastic differential equation.
\end{abstract}

\maketitle

\section{Introduction}

It is almost a tautology that laboratory measurements give rise to
classical stochastic processes.  For example, in quantum optics one
usually detects, using a configuration of photodetectors, the light of a
laser which is scattered off a cloud of atoms, and the resulting
photocurrent is a classical stochastic process\cite{Bar03,BHJ07}.  It is
subsequently of interest to infer as well as possible the state of the
atoms from the observed photocurrent, which is the purpose of quantum
filtering theory.  This theory has been extensively investigated both in
the mathematical literature\cite{Bel92} (see Ref.\ \refcite{BHJ07} for a 
recent review) and in the physics literature, where it is known under the 
name of quantum trajectory theory or the theory of stochastic master 
equations\cite{GaZ04}.

In order to implement the quantum filter, however, the underlying quantum
model is presumed to be known.  It is not evident, \textit{a priori}, that
good estimates will be obtained in the presence of modelling errors which
are inevitable in practice.  Questions of robustness to modelling errors
are particularly subtle on a long time interval, and have received much
attention in the classical nonlinear filtering literature, see, e.g.,
Ref.\ \refcite{CR09} and the references therein.  In particular,
asymptotic stability of the filter---the independence of the filter, after
a long time interval, of the initial estimate of the system---has been
shown to hold in a wide range of classical nonlinear filtering models, and
is the starting point for more general robustness questions.  The problem
of asymptotic stability is related to the consistency of Bayes estimates
and is of significant practical interest as it ensures optimal performance
of the filter, after an initial transient, even under misspecification of
the initial condition.  To date, however, no such result is known in
quantum filtering theory.

The goal of this paper is to develop a criterion which ensures asymptotic
stability of quantum filters.  This \textit{observability} condition for
stability is a natural one: it is the requirement that no two different
initial states of the model give rise to an observation process with the
same law. In the quantum optics example described above, this means that
we must be able to determine precisely the initial state of the atoms if
we have access to the full statistics of the photocurrent over the
infinite time interval.  If this is the case, then the filtered estimates
of the atomic observables are insensitive to the initial state of
the atoms after a long time interval, provided we restrict our 
attention to initial states that satisfy an absolute continuity condition.

The basic method of proof is based on the classical counterpart of this
result, which has recently been developed by the author\cite{Han07}.  To
extend this result to the quantum setting, it is most natural to work
within an abstract quantum filtering setting which is a little more
general than the usual setting in the quantum filtering
literature\cite{Bel92,BHJ07}.  We set up the problem in section
\ref{sec:general} in the context of $C^*$-algebraic Markov process theory
in the spirit of Accardi, Frigerio and Lewis\cite{AFL82}.  The proof of
the main result can be found in section \ref{sec:proofs}.  In section
\ref{sec:abscont} we elaborate on the absolute continuity condition
required by our main result, and provide a simpler sufficient condition.  
In the last section \ref{sec:findim} we investigate a class of quantum
filtering models, defined through the solution of a Hudson-Parthasarathy
type quantum stochastic differential equation with a finite dimensional
initial system, which have important applications, e.g., in quantum
optics.  In this setting one may find explicitly computable rank
conditions for the model to be observable in terms of the coefficients of
the quantum stochastic differential equation and the observation model.

Finally, let us note that the asymptotic stability of nonlinear filters is
not only of interest by itself, but is also an important ingredient in the
development of error bounds for filters under more general modelling
errors or for approximate filters (see, e.g., Refs.\
\refcite{kushner1,CV07} in the classical setting).  In that case, however,
it is typically necessary to obtain more quantitative bounds on the rate
of stability.  Let us also mention that observability, though sufficient,
is not a necessary condition for stability.  One could conjecture that a
natural counterpart of the \textit{detectability} condition in Ref.\
\refcite{Han07} is necessary and sufficient for the stability of quantum
filters in the finite dimensional setting of section \ref{sec:findim}, as
it is in the classical case.

\section{The quantum filtering model}
\label{sec:general}

We will consider quantum filtering theory in the abstract setting of
Feller-type quantum Markov processes in the spirit of Accardi, Frigerio
and Lewis\cite{AFL82,Fag99}.  One of the most important examples in
practice is the quantum stochastic flow generated by a quantum stochastic
differential equation with a finite-dimensional initial system; this
particular setting will be investigated in detail in section
\ref{sec:findim}.  

In this section, we introduce the quantum filtering
model and fix the notation for the rest of the paper.
Let us begin by defining the basic elements of the model.
\begin{itemize}
\item
$\mathcal{A}$, the \textit{initial system}, is a unital $C^*$-algebra
with state space $\mathcal{S}\subset\mathcal{A}^*$;
\item  $\{P_t,t\ge 0\}$ is a one parameter semigroup of contractive and 
completely positive linear maps from $\mathcal{A}$ to itself, with 
$P_0[X]=X$ $\,\forall\,X\in\mathcal{A}$ and $P_t[I]=I$ $\,\forall\,t\ge 0$;
\item $\mathfrak{M}$, the \textit{universal algebra}, is a Von Neumann 
algebra;
\item $\{\mathfrak{M}_{t]}:t\ge 0\}$ is a filtration of subalgebras of 
$\mathfrak{M}$ such that $(\bigcup_{t\ge 
0}\mathfrak{M}_{t]})''=\mathfrak{M}$ and 
$\mathfrak{M}_{0]}\simeq\mathcal{A}^{**}$ (i.e., $\mathfrak{M}_{0]}$ is
the enveloping algebra of $\mathcal{A}$);
\item $\{\Phi_\rho:\rho\in\mathcal{S}\}$ is a family of normal states
on $\mathfrak{M}$ such that the conditional expectations
$\Phi_{\rho}(~\cdot~|\mathfrak{M}_{t]}):\mathfrak{M}\to\mathfrak{M}_{t]}$
exist for every $t\ge 0$ and $\rho\in\mathcal{S}$.
\end{itemize}

\begin{remark}
The requirement that $\mathcal{A}$ be unital is not overly restrictive;
if $\mathcal{A}$ is not unital, we may always enlarge $\mathcal{A}$ by 
adjoining the identity without essentially changing the structure of the 
theory.  When $\mathcal{A}$ is commutative, this corresponds to the 
one-point compactification of the spectrum (Ex.\ VII.8.5 in Ref.\ 
\refcite{Con85}).
\end{remark}

Before proceeding, we recall for the reader's convenience the definition 
of the conditional expectation in a Von Neumann algebra (see, e.g., 
Ref.\ \refcite{OP93}).

\begin{definition}[Conditional expectation]
Let $\mathfrak{A},\mathfrak{A}_0$ be Von Neumann algebras,
$\mathfrak{A}_0\subset\mathfrak{A}$ and let $\Phi$ be a normal state on 
$\mathfrak{A}$.  Suppose there exists a linear map 
$\Phi(~\cdot~|\mathfrak{A}_0):\mathfrak{A}\to\mathfrak{A}_0$ which 
satisfies the following properties:
\begin{itemize}
\item $\Phi(I|\mathfrak{A}_0)=I$;
\item $\Phi(X^*X|\mathfrak{A}_0)\ge 0$ for all $X\in\mathfrak{A}$;
\item $\Phi(X^*|\mathfrak{A}_0)=\Phi(X|\mathfrak{A}_0)^*$ for all
$X\in\mathfrak{A}$;
\item $\Phi(XYZ|\mathfrak{A}_0)=X\Phi(Y|\mathfrak{A}_0)Z$ for all
$Y\in\mathfrak{A}$ and $X,Z\in\mathfrak{A}_0$;
\item $\Phi(\Phi(X|\mathfrak{A}_0)) = \Phi(X)$ for all $X\in\mathfrak{A}$.
\end{itemize}
Then $\Phi(~\cdot~|\mathfrak{A}_0)$ is a \textit{conditional 
expectation} from $\mathfrak{A}$ onto $\mathfrak{A}_0$ with respect to 
$\Phi$.
\end{definition}

It is not difficult to prove that any two maps $P,Q:\mathfrak{A}\to
\mathfrak{A}_0$ which satisfy this definition are
$\Phi$-indistinguishable, i.e., $\Phi(|P(X)-Q(X)|^2)=0$ (see, e.g., Thm.\
3.16 in Ref.\ \refcite{BHJ07}).  Thus the conditional expectation, if it
exists, is essentially unique.  Existence, on the other hand, is not 
guaranteed in the noncommutative setting.

We now return to our filtering setup.  We will presume that there is a 
family $\{j_t:t\ge 0\}$ of $^*$-isomorphisms $j_t:\mathcal{A}\to 
\mathfrak{M}_{t]}$ such that the \textit{Markov property} holds:
$$
	\Phi_\rho(j_{t+s}(X)|\mathfrak{M}_{s]}) =
	j_s(P_t[X])\qquad
	\forall\,t,s\ge 0,~X\in\mathcal{A},~\rho\in\mathcal{S}.
$$
Moreover, we presume that $j_0(\mathcal{A})''=\mathfrak{M}_{0]}$ and that
$$
	\Phi_\rho(j_0(X)) = \rho(X)\qquad
	\forall\,X\in\mathcal{A},~\rho\in\mathcal{S},
$$
i.e., the state $\rho\in\mathcal{S}$ can be interpreted as the 
\textit{initial state} of the quantum Markov process $j_t$.  The latter 
plays the role of the signal process in classical filtering theory.

\begin{remark}
In order that $\Phi_\rho(j_0(X)) = \rho(X)$ for all $\rho$, it is 
necessary that every state $\rho\in\mathcal{S}$ extends to a normal state 
on $\mathfrak{M}_{0]}$.  This forces us to work with the universal 
representation $\mathfrak{M}_{0]}\simeq\mathcal{A}^{**}$ as required 
above, see Thm.\ 1.17.2 in Ref.\ \refcite{Sak98}. 
\end{remark}

In addition, we must introduce the observations. To this end, we introduce
the $n$-dimensional observation process $\{Y_t^k:t\ge 0,~k=1,\ldots,n\}$,
where $Y_t^k$ is a self-adjoint operator affiliated to
$\mathfrak{M}_{t]}$ and $Y_0^k=0$.  Define $\mathfrak{Y}_{t]}$ to be the 
Von Neumann algebra generated by $\{Y_s^k:0\le s\le t,~k=1,\ldots,n\}$. We 
presume that
$$
	\mathfrak{Y}_{t]}\mbox{ is commutative},\qquad
	j_t(X) \in \mathfrak{Y}_{t]}'\qquad
	\forall\,t\ge 0,~X\in\mathcal{A}. 
$$ 
The first condition is known as the \textit{self-nondemolition} property,
and ensures that the process $\{Y_t\}$ can be represented as a classical
stochastic process (as is befitting of laboratory observations). The
second condition is the \textit{nondemolition} property, and ensures that
the conditional expectations $\pi_t^\rho(X) :=
\Phi_\rho(j_t(X)|\mathfrak{Y}_{t]})$ exist for every $X\in\mathcal{A}$ and
$t\ge 0$ (see, e.g., Thm.\ 3.16 in Ref.\ \refcite{BHJ07}).  The 
goal of the
\textit{filtering problem} is to compute these conditional expectations.
This problem can be solved explicitly in specific models, as is known 
since the work of Belavkin\cite{Bel92}; see Ref.\ \refcite{BHJ07} for an 
introduction and review.  For the purpose of this paper, however, it 
will not be necessary to obtain explicit expressions for the filtered 
estimates $\pi_t^\rho(X)$.

Finally, we introduce the following Feller-type assumption.  We presume 
that for any choice of $t_1,\ldots,t_k>0$ and bounded 
continuous functions $f_1,\ldots,f_k:\mathbb{R}\to\mathbb{R}$,
$$
	\Phi_\rho(f_1(Y_{t_1})\cdots 
		f_k(Y_{t_k})|\mathfrak{M}_{0]}) =
	j_0(Z(t_1,\ldots,t_k,f_1,\ldots,f_k))
$$
for some $Z(t_1,\ldots,t_k,f_1,\ldots,f_k)\in\mathcal{A}$ independent of
$\rho$, and moreover
$$
	\Phi_\rho(f_1(Y_{s+t_1}-Y_s)\cdots 
		f_k(Y_{s+t_k}-Y_s)|\mathfrak{M}_{s]}) =
	j_s(Z(t_1,\ldots,t_k,f_1,\ldots,f_k))
$$
for every $s\ge 0$.  The latter assumption ensures, in a sense, that the
observation process is time-homogeneous.  An important example of a
filtering model in which these constructions can be implemented is 
discussed in detail in section \ref{sec:findim}.

The goal of the remainder of the paper is to study the dependence of the 
filter $\pi_t^\rho(X) := \Phi_\rho(j_t(X)|\mathfrak{Y}_{t]})$ on the 
initial state $\rho\in\mathcal{S}$ as $t\to\infty$.

\begin{definition}[Observability]
Let $\mathfrak{Y}= (\bigcup_{t\ge 0}\mathfrak{Y}_{t]})''$.  The model is 
\emph{observable} if there do not exist $\rho_1,\rho_2\in\mathcal{S}$ with 
$\rho_1\ne\rho_2$ and $\Phi_{\rho_1}(Y)=\Phi_{\rho_2}(Y)$ for every 
$Y\in\mathfrak{Y}$.
\end{definition}

We will prove the following result.

\begin{theorem}
\label{thm:main}
If the model is observable, then 
$$
	\Phi_{\rho_1}(
		|\pi^{\rho_1}_t(X)-\pi^{\rho_2}_t(X)|)
	\xrightarrow{t\to\infty}0\qquad
	\forall\,X\in\mathcal{A}
$$
whenever the laws of the observations under $\Phi_{\rho_1}$ and 
$\Phi_{\rho_2}$ are absolutely continuous (i.e., if $P$ is a projection in 
$\mathfrak{Y}$ and $\Phi_{\rho_2}(P)=0$, then $\Phi_{\rho_1}(P)=0$).
\end{theorem}

We can obtain a sufficient condition for the absolute continuity of the
observation laws, as required in theorem \ref{thm:main}, in terms of the
initial states.  This is developed in section \ref{sec:abscont}. In the
finite-dimensional setting, discussed in section \ref{sec:findim}, we will
find explicitly computable conditions for the filtering model to be
observable.

\section{Observability and filter stability}
\label{sec:proofs}

The proof of the main result proceeds in two steps.  First, we establish
that 
$$
	\Phi_{\rho_1}(|\pi^{\rho_1}_t(X)-\pi^{\rho_2}_t(X)|)
	\xrightarrow{t\to\infty} 0
$$ 
for $X$ of the form $Z(t_1,\ldots,t_k,f_1,\ldots,f_k)$.  This holds
without any further assumptions.  Then, we show that the set of all such
observables is total in $\mathcal{A}$ when the model is observable.  A
simple approximation argument then completes the proof.

\subsection{Stability of $Z(t_1,\ldots,t_k,f_1,\ldots,f_k)$}

We begin by proving a simple lemma.  This result is almost trivial---it is
just the tower property of the conditional expectation---but one should
verify that the conditional expectations do in fact exist.

\begin{lemma}
For any $\rho\in\mathcal{S}$, $s\ge 0$ and $t_1,\ldots,t_k,f_1,\ldots,f_k$,
$$
	\pi_s^\rho(Z(t_1,\ldots,t_k,f_1,\ldots,f_k)) = 
	\Phi_\rho(f_1(Y_{s+t_1}-Y_s)\cdots 
		f_k(Y_{s+t_k}-Y_s)|\mathfrak{Y}_{s]})
$$
up to $\Phi$-indistinguishability.
\end{lemma}

\begin{proof}
First, note that by the nondemolition assumption
$$
	\Phi_\rho(f_1(Y_{s+t_1}-Y_s)\cdots 
		f_k(Y_{s+t_k}-Y_s)|\mathfrak{M}_{s]}) =
	j_s(Z(t_1,\ldots,t_k,f_1,\ldots,f_k)) \in \mathfrak{Y}_{s]}'.
$$
Hence the conditional expectation with respect to $\mathfrak{Y}_{s]}$ 
exists and
$$
	\pi_s^\rho(Z(t_1,\ldots,t_k,f_1,\ldots,f_k)) = 
	\Phi_\rho(\Phi_\rho(f_1(Y_{s+t_1}-Y_s)\cdots 
		f_k(Y_{s+t_k}-Y_s)|\mathfrak{M}_{s]})|\mathfrak{Y}_{s]}).
$$
Moreover, the conditional expectation
$$
	\Phi_\rho(f_1(Y_{s+t_1}-Y_s)\cdots 
		f_k(Y_{s+t_k}-Y_s)|\mathfrak{Y}_{s]})
$$
exists as $f_1(Y_{s+t_1}-Y_s)\cdots
f_k(Y_{s+t_k}-Y_s)\in\mathfrak{Y}_{s]}'$ (this follows directly as
$\mathfrak{Y}$ is commutative).  Finally, note that observables of the
form $X\,f_1(Y_{s+t_1}-Y_s)\cdots f_k(Y_{s+t_k}-Y_s)$ with
$X\in\mathfrak{Y}_{s]}$ are weak$^*$ total in $\mathfrak{Y}$. Hence the
maps $\Phi_\rho(~\cdot~|\mathfrak{Y}_{s]}):\mathfrak{Y}\to\mathfrak{Y}_{s]}$
and $\Phi_\rho(\Phi_\rho(~\cdot~|\mathfrak{M}_{s]})|
\mathfrak{Y}_{s]}):\mathfrak{Y}\to\mathfrak{Y}_{s]}$ are both well
defined.  It remains to note that both these maps satisfy the definition
of the conditional expectation. 
\end{proof}

We can now prove the stability of $Z(t_1,\ldots,t_k,f_1,\ldots,f_k)$.  By 
virtue of the previous lemma the setting is essentially classical (as all 
the objects involved live in the commutative algebra $\mathfrak{Y}$), and 
we will exploit this fact explicitly in the proof.

\begin{proposition}
\label{prop:exclint}
Suppose that the law of the observations under $\Phi_{\rho_1}$ is 
absolutely continuous with respect to the law of the observations under 
$\Phi_{\rho_2}$.  Then
$$
	\Phi_{\rho_1}(|\pi^{\rho_1}_t(Z(t_1,\ldots,t_k,f_1,\ldots,f_k))
		-\pi^{\rho_2}_t(Z(t_1,\ldots,t_k,f_1,\ldots,f_k))|)
	\xrightarrow{t\to\infty} 0
$$ 
for any $t_1,\ldots,t_k>0$ and bounded continuous 
functions $f_1,\ldots,f_k$.
\end{proposition}

\begin{proof}
We work exclusively on the commutative algebra $\mathfrak{Y}$.  By the
spectral theorem (Prop.\ 1.18.1 in Ref.\ \refcite{Sak98}), 
there exists a measure
space $(\Omega,\mathscr{F},\lambda)$ which admits a surjective
$^*$-isomorphism $\iota:\mathfrak{Y}\to
L^\infty(\Omega,\mathscr{F},\lambda)$, and every state $\Phi_\varphi$
induces a probability measure $\mathbf{P}_\varphi$ on $\Omega$ such that
$\Phi_\varphi(X) = \mathbf{E}_\varphi(\iota(X))$ for all
$X\in\mathfrak{Y}$.  Moreover, there exists a classical stochastic process
$\{y_t^k:t\ge 0,~k=1,\ldots,n\}$ on $\Omega$ such that
$$
	\iota(f(Y_{t_1}^{k_1},\ldots,Y_{t_\ell}^{k_\ell}))=
	f(y_{t_1}^{k_1},\ldots,y_{t_\ell}^{k_\ell})
	\qquad\forall\mbox{ bounded measurable }
	f:\mathbb{R}^\ell\to\mathbb{R},
$$ 
and it is straightforward to verify that 
\begin{multline*}
	\iota(\Phi_\varphi(f_1(Y_{s+t_1}-Y_s)\cdots 
	f_k(Y_{s+t_k}-Y_s)|\mathfrak{Y}_{s]})) \\
	= \mathbf{E}_\varphi(
	f_1(y_{s+t_1}-y_s)\cdots f_k(y_{s+t_k}-y_s)|\mathscr{Y}_{s})
\end{multline*}
where $\mathscr{Y}_s=\sigma\{y_r:0\le r\le s\}$.
Evidently it suffices to prove that
$$
	\mathbf{E}_{\rho_1}(|
		\mathbf{E}_{\rho_1}(
		\xi_t|\mathscr{Y}_{t})
		- \mathbf{E}_{\rho_2}(\xi_t|\mathscr{Y}_{t})
	|)\xrightarrow{t\to\infty} 0
$$
whenever $\xi_s=f_1(y_{s+t_1}-y_s)\cdots f_k(y_{s+t_k}-y_s)$ for all $s\ge 
0$ and $\rho_1,\rho_2\in\mathcal{S}$ which give rise to absolutely continuous
observation laws.

To proceed, note that by our absolute continuity assumption
$\mathbf{P}_{\rho_1}|_{\mathscr{Y}_\infty}\ll
\mathbf{P}_{\rho_2}|_{\mathscr{Y}_\infty}$.  We can therefore apply the 
classical Bayes formula (Lem.\ 8.6.2 in Ref.\ \refcite{Oks98}):
$$
	\mathbf{E}_{\rho_2}(\Delta|\mathscr{Y}_{t})\,
	\mathbf{E}_{\rho_1}(\xi_t|\mathscr{Y}_{t}) =
	\mathbf{E}_{\rho_2}(\Delta\,\xi_t|\mathscr{Y}_{t})\qquad
	\mathbf{P}_{\rho_2}\mbox{-a.s.},
$$
where $\Delta = d\mathbf{P}_{\rho_1}|_{\mathscr{Y}_\infty}/
d\mathbf{P}_{\rho_2}|_{\mathscr{Y}_\infty}$ 
($\mathscr{Y}_\infty=\bigvee_{t\ge 0}\mathscr{Y}_t$).  Thus we find that
$$
	\mathbf{E}_{\rho_2}(\Delta|\mathscr{Y}_{t})\,|
	\mathbf{E}_{\rho_1}(\xi_t|\mathscr{Y}_{t}) - 
	\mathbf{E}_{\rho_2}(\xi_t|\mathscr{Y}_{t})| =
	|\mathbf{E}_{\rho_2}((\Delta-\mathbf{E}_{\rho_2}
	(\Delta|\mathscr{Y}_{t}))\,\xi_t|\mathscr{Y}_{t})|
	\qquad
	\mathbf{P}_{\rho_2}\mbox{-a.s.}
$$
Taking the expectation with respect to $\mathbf{P}_{\rho_2}$, we obtain
$$
	\mathbf{E}_{\rho_1}(|
	\mathbf{E}_{\rho_1}(\xi_t|\mathscr{Y}_{t}) - 
	\mathbf{E}_{\rho_2}(\xi_t|\mathscr{Y}_{t})|) =
	\mathbf{E}_{\rho_2}(|\mathbf{E}_{\rho_2}((\Delta-\mathbf{
	E}_{\rho_2}(\Delta|\mathscr{Y}_{t}))\,\xi_t|\mathscr{Y}_{t})|).
$$
By Jensen's inequality
$$
	\mathbf{E}_{\rho_2}(|\mathbf{E}_{\rho_2}((\Delta-\mathbf{
	E}_{\rho_2}(\Delta|\mathscr{Y}_{t}))\,\xi_t|\mathscr{Y}_{t})|)
	\le
	K\,\mathbf{E}_{\rho_2}(
	|\Delta-\mathbf{E}_{\rho_2}(\Delta|\mathscr{Y}_{t})|),
$$
where $K=\|f_1\|_\infty\cdots\|f_k\|_\infty$.  But note that $\xi_t$ is 
measurable with respect to $\mathscr{Y}_\infty$, so by the martingale
convergence theorem $\mathbf{E}_{\rho_2}(\Delta|\mathscr{Y}_{t}) \to 
\Delta$ in $L^1(\Omega,\mathscr{F},\mathbf{P}_{\rho_2})$.
Therefore $\mathbf{E}_{\rho_1}(|
\mathbf{E}_{\rho_1}(\xi_t|\mathscr{Y}_{t}) - 
\mathbf{E}_{\rho_2}(\xi_t|\mathscr{Y}_{t})|) \to 0$, and the proof is 
complete.
\end{proof}

\begin{corollary}
\label{cor:cordense}
Denote by $\mathcal{O}^0\subset\mathcal{A}$ the linear span of 
$Z(t_1,\ldots,t_k,f_1,\ldots,f_k)$ for all 
$t_1,\ldots,t_k,f_1,\ldots,f_k$, and suppose that the law of the 
observations under $\Phi_{\rho_1}$ is absolutely continuous with respect 
to the law of the observations under $\Phi_{\rho_2}$.  Then
$$
	\Phi_{\rho_1}(|\pi^{\rho_1}_t(Z)-\pi^{\rho_2}_t(Z)|)
	\xrightarrow{t\to\infty} 0\qquad
	\forall\,Z\in\mathrm{cl}\,\mathcal{O}^0,
$$ 
where $\mathrm{cl}\,\mathcal{O}^0$ denotes the (uniform) closure of 
$\mathcal{O}^0$ in $\mathcal{A}$.
\end{corollary}

\begin{proof}
Fix $Z\in\mathrm{cl}\,\mathcal{O}^0$ and a sequence
$\{Z_n\}\subset\mathcal{O}^0$ such that $\|Z_n-Z\|\to 0$ as $n\to\infty$.
For every $n<\infty$, we have $\Phi_{\rho_1}(|\pi^{\rho_1}_t(Z_n)-
\pi^{\rho_2}_t(Z_n)|)\to 0$ as $t\to\infty$; to see this, it suffices to 
use the linearity of the conditional expectation and the fact that the 
triangle inequality holds for $|\,\cdot\,|$ when we restrict our attention 
to a commutative algebra (i.e., $|\sum_i X_i|\le\sum_i|X_i|$ provided that 
the $X_i$ commute with each other and their adjoints).  Reasoning in the 
same way, we find immediately that
\begin{multline*}
	\Phi_{\rho_1}(|\pi^{\rho_1}_t(Z)-\pi^{\rho_2}_t(Z)|) \\ \le
	\Phi_{\rho_1}(|\pi^{\rho_1}_t(Z-Z_n)|) + 
	\Phi_{\rho_1}(|\pi^{\rho_1}_t(Z_n)-\pi^{\rho_2}_t(Z_n)|) +
	\Phi_{\rho_1}(|\pi^{\rho_2}_t(Z_n-Z)|).
\end{multline*}
The first and the third term on the right are bounded above by
$\|Z_n-Z\|$.  Hence
$$
	\limsup_{t\to\infty}
	\Phi_{\rho_1}(|\pi^{\rho_1}_t(Z)-\pi^{\rho_2}_t(Z)|) \le 
	2\,\|Z_n-Z\|.
$$
The result follows by letting $n\to\infty$.
\end{proof}

\subsection{Observability and approximation}

From the previous corollary, we see that a sufficient condition for the
stability of the filter is that $\mathrm{cl}\,\mathcal{O}^0 =
\mathcal{A}$.  We will show that this is the case if and only if the model
is observable.  In fact, we will prove a more general result, from which
this statement follows.  We begin with the following definition.

\begin{definition}[Observable space]
For $\rho_1,\rho_2\in\mathcal{S}$, we define the equivalence relation 
$\rho_1\backsim\rho_2$ whenever $\Phi_{\rho_1}(Y)=\Phi_{\rho_2}(Y)$ for 
every $Y\in\mathfrak{Y}$.  The Banach space
$$
	\mathcal{O} = \{X\in\mathcal{A}:
		\rho_1(X)=\rho_2(X)\mbox{ for all }
		\rho_1,\rho_2\in\mathcal{S}\mbox{ such that }
		\rho_1\backsim\rho_2\}
$$
is called the observable space of the model.
\end{definition}

The following result is key.

\begin{proposition}
\label{prop:density}
$\mathcal{O}^0$ is dense in $\mathcal{O}$.
\end{proposition}

\begin{proof}
Suppose that $\mathcal{O}^0$ is not dense in $\mathcal{O}$.  Then there 
must be an element $X$ of $\mathcal{O}$ that is not in 
$\mathrm{cl}\,\mathcal{O}^0$.  By the Hahn-Banach theorem, there exists 
an element $\varphi\in\mathcal{A}^*$ such that $\varphi(Z)=0$ for all
$Z\in\mathrm{cl}\,\mathcal{O}^0$ and $\varphi(X)\ne 0$.  Then either
$\varphi(X)+\varphi(X)^*\ne 0$, or $i(\varphi(X)-\varphi(X)^*)\ne 0$, so
we may assume without loss of generality that $\varphi$ is real-valued.
In particular, we can write $\varphi=\varphi_1-\varphi_2$ where
$\varphi_1,\varphi_2$ are nonnegative (e.g., Prop.\ 1.17.1 in 
Ref.\ \refcite{Sak98}).  But note that $I\in\mathcal{O}^0$, so
$\varphi_1(I)=\varphi_2(I)$.  We can thus define 
$\rho_1,\rho_2\in\mathcal{S}$ by $\rho_1=\varphi_1/
\varphi_1(I)$ and $\rho_2=\varphi_2/\varphi_2(I)$, and we find that
$\rho_1(X)\ne\rho_2(X)$ and $\rho_1(Z)=\rho_2(Z)$ for all 
$Z\in\mathrm{cl}\,\mathcal{O}^0$.  Now note that for any 
$\rho\in\mathcal{S}$
$$
	\rho(Z(t_1,\ldots,t_k,f_1,\ldots,f_k)) =
	\Phi_{\rho}(f_1(Y_{t_1})\cdots f_k(Y_{t_k})).
$$
Hence we find that
$$
	\Phi_{\rho_1}(f_1(Y_{t_1})\cdots f_k(Y_{t_k})) = 
	\Phi_{\rho_2}(f_1(Y_{t_1})\cdots f_k(Y_{t_k}))
$$
for all $t_1,\ldots,t_k,f_1,\ldots,f_k$.  As the set of observables of the 
form $f_1(Y_{t_1})\cdots f_k(Y_{t_k})$ is weak$^*$ total in 
$\mathfrak{Y}$, we conclude that $\Phi_{\rho_1}(Y)=\Phi_{\rho_2}(Y)$
for all $Y\in\mathfrak{Y}$.  But then $\rho_1\backsim\rho_2$, which 
implies $\rho_1(X)=\rho_2(X)$, and we have a contradiction.
\end{proof}

We immediately find the following corollary.

\begin{corollary}
\label{cor:maingen}
Suppose that the law of the observations under $\Phi_{\rho_1}$ is
absolutely continuous with respect to the law of the observations under
$\Phi_{\rho_2}$.  Then
$$
	\Phi_{\rho_1}(
		|\pi^{\rho_1}_t(X)-\pi^{\rho_2}_t(X)|)
	\xrightarrow{t\to\infty}0\qquad
	\forall\,X\in\mathcal{O}.
$$
\end{corollary}

\begin{proof}
Immediate from corollary \ref{cor:cordense} and 
$\mathrm{cl}\,\mathcal{O}^0=\mathcal{O}$.
\end{proof}

We may finally complete the proof of theorem \ref{thm:main}.

\begin{proof} (Theorem \ref{thm:main}).
The model is observable, by definition, if $\rho_1\backsim\rho_2$ 
implies $\rho_1=\rho_2$.  Clearly this is the case if and only if
$\mathcal{O}=\mathcal{A}$.  The result follows directly.
\end{proof}

\begin{remark}
The proof of proposition \ref{prop:density} clarifies why it is important 
to work in the $C^*$-algebraic setting, rather than starting off with an 
initial Von Neumann algebra.  As the state space of a $C^*$-algebra is 
dual to the algebra itself, we may employ the Hahn-Banach theorem as in 
the proof of proposition \ref{prop:density} to characterize the observable 
space.  For a Von Neumann algebra, however, the space of normal states is 
predual to the algebra.  To employ the technique used in the proof of 
proposition \ref{prop:density}, we would then have two options: we must 
either consider non-normal initial states, or prove density of 
$\mathcal{O}^0$ in $\mathcal{O}$ in the weak$^*$ topology on the initial 
Von Neumann algebra.  The former is unphysical, while in the latter case 
corollary \ref{cor:cordense} can not be employed.  It thus appears that 
the $C^*$-algebraic setting is the natural setting in which our results 
can be developed.
\end{remark}

\section{Absolute continuity and randomization}
\label{sec:abscont}

In our main result, theorem \ref{thm:main}, we required that the initial
state $\rho_1,\rho_2\in\mathcal{S}$ are such that the law of the
observations under $\Phi_{\rho_1}$ is absolutely continuous with respect
to the law of the observations under $\Phi_{\rho_2}$.  One might expect
that a sufficient condition would be that the initial states are
themselves absolutely continuous in a suitable sense.  The goal of this
section is to develop this idea.

Before we turn to the filtering model of section \ref{sec:general}, let us 
consider the general setting where $\mathcal{A}$ is any unital 
$C^*$-algebra.  Given a state $\varphi$ on $\mathcal{A}$, we denote by
$(\pi_\varphi,\mathsf{H}_\varphi,\xi_\varphi)$ the cyclic representation
of $\mathcal{A}$ induced by $\varphi$.

Let $\mathcal{S}\subset\mathcal{A}^*$ denote the state space of
$\mathcal{A}$.  We endow $\mathcal{A}^*$ with the weak$^*$ topology, and
recall that this makes $\mathcal{S}$ a compact convex set. By a (finite)
measure on $\mathcal{S}$ we mean a regular Borel measure on $\mathcal{S}$
or, equivalently, an element of $C(\mathcal{S})^*$ (see p.\ 232 
in Ref.\ \refcite{Tak02}).  A probability measure is a 
nonnegative measure with unit mass.

We now recall a basic construction in Choquet theory.  Let $\mu$ be
a probability measure on $\mathcal{S}$.  Then (Lem.\ IV.6.3 in Ref.\ 
\refcite{Tak02})  there is a unique $\rho\in\mathcal{S}$ such that
$$
	F(\rho) = \int_{\mathcal{S}} F(\varphi)\,\mu(d\varphi)\qquad
	\forall\,F\in\mathcal{A}^{**}.
$$
The state $\rho$ is called the \textit{barycenter} of the probability
measure $\mu$.  The measure $\mu$ can be thought of as a 
\textit{randomization} of the state $\rho$; indeed, we have replaced the 
state $\rho$ by a random state, with law $\mu$, which averages to $\rho$:
$$
	\rho(X) = \int_{\mathcal{S}} \varphi(X)\,\mu(d\varphi)\qquad
	\forall\,X\in\mathcal{A}.
$$
The idea is now to seek randomizations which have desirable probabilistic
properties.  In particular, we will consider the following notion of 
absolute continuity.

\begin{definition}[Absolute continuity]
\label{defn:abscont}
The state $\rho_1\in\mathcal{S}$ is \emph{absolutely 
continuous} with respect to $\rho_2\in\mathcal{S}$, denoted as 
$\rho_1\ll\rho_2$, if there exist probability measures $\mu_1,\mu_2$ on 
$\mathcal{S}$ such that $\rho_1$ is the barycenter of $\mu_1$, $\rho_2$ is 
the barycenter of $\mu_2$, and $\mu_1\ll\mu_2$.
\end{definition}

We now show that this natural definition of absolute continuity of
$\rho_1$ with respect to $\rho_2$ is equivalent to the requirement that
$\rho_1$ is \textit{presque domin{\'e}e} (almost dominated) by $\rho_2$ in
the sense of Dixmier (Ch.\ I, \S 4, Ex.\ 8c in Ref.\ \refcite{Dix69}).
Radon-Nikodym type results in this setting have been investigated by
Naudts\cite{Nau74} and Gudder\cite{Gud79}.

\begin{proposition}
\label{prop:abscont}
Let $\rho_1,\rho_2\in\mathcal{S}$.  Then the following
are equivalent:
\begin{enumerate}
\item $\rho_1\ll\rho_2$;
\item For every sequence $\{X_n\}\subset\mathcal{A}$ such that
$\lim_{m,n}\rho_1((X_m-X_n)^*(X_m-X_n))=0$, we have 
$\lim_n\rho_1(X_n^*X_n)=0$ whenever $\lim_n\rho_2(X_n^*X_n)=0$.
\item There exists a positive self-adjoint operator $T$ on 
$\mathsf{H}_{\rho_2}$, affiliated to $\pi_{\rho_2}(\mathcal{A})'$, such 
that $\rho_1(X)=\langle T\xi_{\rho_2},
\pi_{\rho_2}(X)T\xi_{\rho_2}\rangle$ for all $X\in\mathcal{A}$.
\end{enumerate}
\end{proposition}

\begin{proof}

\textbf{($1\Rightarrow 2$)} As $\rho_1\ll\rho_2$, there are probability 
measures $\mu,\nu$ on $\mathcal{S}$ with $\mu\ll\nu$ and
$$
	\rho_1(X) = \int_{\mathcal{S}} 
		\varphi(X)\,\mu(d\varphi),\qquad
	\rho_2(X) = \int_{\mathcal{S}}\varphi(X)\,\nu(d\varphi),\qquad
	\forall\,X\in\mathcal{A}.
$$
Let $\{X_n\}$ be such that $\lim_{m,n}\rho_1((X_m-X_n)^*(X_m-X_n))=0$, and
define the random variables $\Phi_n:\mathcal{S}\to[0,\infty[\mbox{}$ by
$\Phi_n(\varphi) = \varphi(X_n^*X_n)$.  We begin by establishing that
$\{\Phi_n\}$ is a Cauchy sequence in $L^1(\mathcal{S},\mu)$. 
By Lem.\ 2.4 in Ref.\ \refcite{Nie83}
$$
	|\Phi_m(\varphi)-\Phi_n(\varphi)|
	\le
	\varphi((X_m-X_n)^*(X_m-X_n))^{1/2}
	\left[
		\varphi(X_m^*X_m)^{1/2}
		+ \varphi(X_n^*X_n)^{1/2}
	\right].
$$
Therefore, we find using the Cauchy-Schwarz inequality and
$(a+b)^2\le 2a^2+2b^2$
$$
	\left[
	\int_{\mathcal{S}} |\Phi_m-\Phi_n|\,d\mu
	\right]^2 \le
	2\rho_1(X_m^*X_m+X_n^*X_n)\,
	\rho_1((X_m-X_n)^*(X_m-X_n)).
$$
Thus $\{\Phi_n\}$ is Cauchy in $L^1(\mathcal{S},\mu)$ provided that
$\rho_1(X_n^*X_n)$ converges to a finite limit. To show this, define
$\psi_n\in\mathsf{H}_{\rho_1}$ by $\psi_n=\pi_{\rho_1}(X_n)\xi_{\rho_1}$.  
Then $\rho_1(X_n^*X_n)=\|\psi_n\|^2$ and
$\rho_1((X_m-X_n)^*(X_m-X_n))=\|\psi_m-\psi_n\|^2$. As the latter
converges to zero, we see that $\{\psi_n\}$ is a Cauchy sequence in
$\mathsf{H}_{\rho_1}$ and thus $\rho_1(X_n^*X_n)$ has a finite limit.

Now suppose that, in addition, $\lim_{n}\rho_2(X_n^*X_n)=0$. Then
evidently $\Phi_n\to 0$ in $L^1(\mathcal{S},\nu)$, so that in particular
$\Phi_n\to 0$ in $\nu$-probability as well as in $\mu$-probability (as
$\mu\ll\nu$).  But as $\{\Phi_n\}$ is a Cauchy sequence in
$L^1(\mathcal{S},\mu)$, it follows that $\Phi_n\to 0$ in
$L^1(\mathcal{S},\mu)$.  Thus $\lim_n\rho_1(X_n^*X_n)=0$, which is what we
set out to prove. 

\textbf{($2\Leftrightarrow 3$)} See Cor.\ 2 in Ref.\ \refcite{Gud79}.

\textbf{($3\Rightarrow 1$)} Denote by $\mathfrak{C}$ the commutative Von 
Neumann algebra generated by $T$ (i.e., this is the smallest Von Neumann 
subalgebra of $\mathfrak{B}(\mathsf{H}_{\rho_2})$ which contains the 
spectral projections of $T$).  By \S 3.1 in Ref.\ \refcite{Sak98}, there 
is a unique probability measure $\nu$ on $\mathcal{S}$ with barycenter 
$\rho_2$ and surjective $^*$-isomorphism $\Gamma:\mathfrak{C}\to 
L^\infty(\mathcal{S},\nu)$ so that
$$
	\langle\xi_{\rho_2},\pi_{\rho_2}(X)V\xi_{\rho_2}\rangle
	= \int_{\mathcal{S}} \Gamma(V)(\varphi)\,\varphi(X)\,
		\nu(d\varphi)\qquad
	\forall\,V\in\mathfrak{C},~X\in\mathcal{A}.
$$
Now define $f_n(x) = nx/(n+x)$ and set $T_n = f_n(T)$.  Then 
$T_n\in\mathfrak{C}$ is a bounded, self-adjoint operator and,
writing the spectral measure of $T$ as $E_T(d\lambda)$, we find that
$$
	\|T_n\xi_{\rho_2}-T\xi_{\rho_2}\|^2 =
	\int_{[0,\infty[}
		|f_n(\lambda)-\lambda|^2\,
	\langle\xi_{\rho_2},E_T(d\lambda)\xi_{\rho_2}\rangle
	\xrightarrow{n\to\infty}0
$$
by dominated convergence (as $|f_n(\lambda)-\lambda|^2\le
2f_n(\lambda)^2+2\lambda^2\le 4\lambda^2$, and 
$\|T\xi_{\rho_2}\|^2<\infty$ by construction).  Consequently, we obtain
$$
	\int_{\mathcal{S}} \Gamma(T_n^2)(\varphi)\,\varphi(X)\,
		\nu(d\varphi)
	=
	\langle T_n\xi_{\rho_2},\pi_{\rho_2}(X)T_n\xi_{\rho_2}\rangle
	\xrightarrow{n\to\infty}\rho_1(X).
$$
But note that $\Gamma(T_n^2)(\varphi)$ is nonnegative, nondecreasing and
$$
	\int_{\mathcal{S}} \Gamma(T_n^2)(\varphi)\,
		\nu(d\varphi)
	\xrightarrow{n\to\infty}1,
$$
so by monotone convergence $\Gamma(T_n^2)\nearrow\Delta$ with $\Delta\in 
L^1(\mathcal{S},\nu)$.  Thus
$$
	\rho_1(X) = 
	\int_{\mathcal{S}} \Delta(\varphi)\,\varphi(X)\,
		\nu(d\varphi)\qquad\forall\, X\in\mathcal{A}
$$
by dominated convergence.  We now define $d\mu=\Delta\,d\nu$, and $\mu$ 
has barycenter $\rho_1$.
\end{proof}

We now return to the filtering setting of section \ref{sec:general}.
The following result establishes that absolute continuity of the initial 
states is indeed a sufficient condition for absolute continuity of the 
observation laws.  Absolute continuity of the initial states is often 
easily verified, e.g., in the finite dimensional setting of section 
\ref{sec:findim}.

\begin{proposition}
\label{prop:ac}
Suppose that $\rho_1\ll\rho_2$.  Then the law of the observations under 
$\Phi_{\rho_1}$ is absolutely continuous with respect to the law of the 
observations under $\Phi_{\rho_2}$.
\end{proposition}

\begin{proof}
As in the proof of proposition \ref{prop:exclint}, we begin by 
constructing a measure space $(\Omega,\mathscr{F},\lambda)$, a 
surjective $^*$-isomorphism $\iota:\mathfrak{Y}\to L^\infty(\Omega,
\mathscr{F},\lambda)$, and a family of probability measures 
$\mathbf{P}_\varphi$ on $\Omega$ such that
$\Phi_\varphi(X) = \mathbf{E}_\varphi(\iota(X))$ for all
$X\in\mathfrak{Y}$.

As $\rho_1\ll\rho_2$, there exist two probability measures $\mu_1,\mu_2$ 
such that $\rho_1$ is the barycenter of $\mu_1$, $\rho_2$ is the 
barycenter of $\mu_2$, and $\mu_1\ll\mu_2$.  We will utilize these 
measures to construct randomizations of the classical probability measures 
$\mathbf{P}_{\rho_1}$ and $\mathbf{P}_{\rho_2}$.  To this end, let 
$\mathscr{B}$ be the Borel $\sigma$-algebra on the state space 
$\mathcal{S}$, and construct the enlarged probability space 
$(\tilde\Omega,\mathscr{\tilde F},\mathbf{\tilde P}_2)$ by setting
$\tilde\Omega=\Omega\times\mathcal{S}$,
$\mathscr{\tilde F}=\mathscr{F}\times\mathscr{B}$, and
$$
        \mathbf{\tilde P}_2(A)=
                \int_{\mathcal{S}}\left[
                \int_{\Omega}
                        I_A(\omega,\varphi)\,
                        \mathbf{P}_{\varphi}(d\omega)
                \right]\mu_2(d\varphi) \qquad
        \forall\,A\in\mathscr{\tilde F}.
$$
Moreover, we define $d\mathbf{\tilde 
P}_1=\Delta\,d\mathbf{\tilde P}_2$
where $\Delta(\omega,\varphi) = (d\mu_1/d\mu_2)(\varphi)$.  
Then
$$
        \int_{\tilde\Omega} F(\omega)\,\mathbf{\tilde P}_i(d\omega,d\varphi)
        = \int_{\Omega} F(\omega)\,\mathbf{P}_{\rho_i}(d\omega),\qquad
        i=1,2,
$$
for all bounded measurable functions $F:\Omega\to\mathbb{R}$.  As by 
construction $\mathbf{\tilde P}_1\ll\mathbf{\tilde P}_2$, clearly the 
marginals satisfy $\mathbf{P}_{\rho_1}\ll\mathbf{P}_{\rho_2}$ also.  The 
proof is complete.
\end{proof}

\section{The finite dimensional case}
\label{sec:findim}

In this section, we consider a specific class of quantum filtering models
which have important applications in quantum optics (see, e.g., Refs.\
\refcite{Bar03,BHJ07}).

Fix $p,q\in\mathbb{N}$ and let
$\mathsf{H}=\mathbb{C}^p\otimes\mathsf{\Gamma}$, where
$\mathsf{\Gamma}=\mathsf{\Gamma_s}(L^2(\mathbb{R}_+)\otimes\mathbb{C}^q)$
is the symmetric Fock space of multiplicity $q$. Thus $p$ is the dimension
of the initial system, while $q$ is the noise dimension.  We set
$\mathcal{A} = M_p$ (the $^*$-algebra of $p\times p$ complex matrices),
$\mathfrak{M}=\mathfrak{B}(\mathsf{H})=\mathcal{A}\otimes
\mathfrak{B}(\mathsf{\Gamma})$.  Moreover, recalling that the Fock space
admits the natural tensor product structure
$\mathsf{\Gamma}=\mathsf{\Gamma}_{t]}\otimes\mathsf{\Gamma}_{[t}$, we
define the filtration of subalgebras
$$
	\mathfrak{M}_{t]}=\{X\otimes I:
	X\in\mathcal{A}\otimes\mathfrak{B}(\mathsf{\Gamma}_{t]})\}.
$$
Finally, we define the family of states $\Phi_\rho = \rho\otimes\Phi_V$ 
with $\Phi_V(X)=\langle\xi,X\xi\rangle$, where $\xi$ is the vacuum vector 
in $\mathsf{\Gamma}$.  It is not difficult to verify that the conditional
expectations $\Phi_\rho(~\cdot~|\mathfrak{M}_{t]})$ exist in this setting;  
in fact, they are given explicitly as follows:
$$
	H_{t]}:\mathbb{C}^p\otimes\mathsf{\Gamma}_{t]}\to
	\mathbb{C}^p\otimes\mathsf{\Gamma},\quad
	H_{t]}\psi := \psi\otimes\xi_{[t},\qquad\quad
	\Phi_\rho(X|\mathfrak{M}_{t]}) = H_{t]}^*XH_{t]}\otimes I,
$$
where $\xi_{[t}$ is the vacuum vector in $\mathsf{\Gamma}_{[t}$.  See, 
e.g., Ref.\ \refcite{Par92} for further details.

We now introduce, as usual, the canonical quantum noises $A_i(t),
A_i^\dag(t),\Lambda_{ij}(t)$, $i,j=1,\ldots,q$ on $\mathsf{\Gamma}$ (we 
will denote their ampliations to $\mathsf{H}$ by the same notation), and 
consider the Hudson-Parthasarathy quantum stochastic differential equation
\begin{multline*}
	dU_t = 
	\Bigg\{\sum_{i,j=1}^{q_0}(S_{ij}-\delta_{ij})\,d\Lambda_{ij}(t)
		+ \sum_{i=1}^{q_0} L_i\,dA_i^\dag(t)
	\\
		- \sum_{i,k=1}^{q_0} L_k^*S_{ki}\,dA_i(t)
		- \frac{1}{2}\sum_{k=1}^{q_0} L_k^*L_k\,dt
		-iH\,dt
	\Bigg\}\,U_t,\qquad
	U_0=I,
\end{multline*}
where $q_0\le q$ and $S_{ij},L_i,H\in\mathcal{A}$, $H$ is self-adjoint,
and $\sum_{ij}S_{ij}\otimes e_ie_j^*$ is a unitary operator in
$M_p\otimes M_q$ ($e_i$ is the $i$th basis vector in $\mathbb{C}^q$).  
Then this equation has a unique solution $\{U_t:t\ge 0\}$ such that
$U_t$ is unitary for every $t\ge 0$ (Thm.\ 27.8 in Ref.\ \refcite{Par92}).  
Moreover, if we define $j_t:\mathcal{A}\to\mathfrak{M}_{t]}$ by $j_t(X) = 
U_t^*(X\otimes I)U_t$, then $j_t$ satisfies the quantum Markov property 
for the semigroup $\{P_t:t\ge 0\}$ generated by
$$
	\mathscr{L}[X] = \lim_{t\searrow 0}\frac{P_t[X]-X}{t} =
	i[H,X] + \sum_{k=1}^{q_0}\left\{L_k^*XL_k
		-\frac{1}{2} L_k^*L_kX-\frac{1}{2} XL_k^*L_k
	\right\},
$$
see Cor.\ 27.10 in Ref.\ \refcite{Par92}.  As by construction
$\Phi_\rho(j_0(X)) = \rho(X)$ for any $\rho\in\mathcal{S}$, this
model satisfies the requirements of section \ref{sec:general}.  

It remains to introduce the observations.  For sake of concreteness, we
will consider in detail two common observation models: a one-dimensional
homodyne detection model and a one-dimensional photon counting model.  
The generalization of these results to other observation models and to
higher dimensional observations is straightforward.  Before proceeding,
however, we prove the following simple lemma.

\begin{lemma}
Let $\rho_1,\rho_2$ be states on $M_p$ which are defined by the density 
matrices $\varrho_1,\varrho_2$ (i.e., $\rho_i(X) = \mathrm{Tr}[\varrho_i
X]$).  Then $\rho_1\ll\rho_2$ if and only if $\ker\varrho_1\supset
\ker\varrho_2$.
\end{lemma}

\begin{proof}
We first prove that $\ker\varrho_1\supset\ker\varrho_2$ implies
$\rho_1\ll\rho_2$.  Let us restrict $\varrho_1,\varrho_2$ to the
subspace $\mathsf{h}=(\ker\varrho_2)^\perp$.  Note that
$\varrho_2|_{\mathsf{h}}$ has full rank and hence is positive definite, so
there is some $\varepsilon\in\mbox{}]0,1[\mbox{}$ such that $\langle
v,\varrho_2v\rangle\ge\varepsilon\|v\|^2$ for all $v\in\mathsf{h}$.  But
the eigenvalues of $\varrho_1$ must be contained in $[0,1]$, so that
$\langle v,\varrho_1v\rangle\le \|v\|^2$ for any $v\in\mathsf{h}$.  Thus
we find that $\langle v,\varrho_2v\rangle\ge\varepsilon\langle v,\varrho_1
v\rangle$ for all $v\in\mathsf{h}$, so evidently 
$\varrho_2\ge\varepsilon\varrho_1$.  But then
$\varrho_1'=(\varrho_2-\varepsilon\varrho_1)/(1-\varepsilon)$ defines 
another state $\rho_1'$ on $M_p$, and the measures $\mu_1\ll\mu_2$ where
$\mu_1 = \delta_{\{\rho_1\}}$ and $\mu_2 = 
\varepsilon\delta_{\{\rho_1\}} + (1-\varepsilon)\delta_{\{\rho_1'\}}$
have barycenters $\rho_1$ and $\rho_2$.

It remains to prove that $\rho_1\ll\rho_2$ implies
$\ker\varrho_1\supset\ker\varrho_2$.  To this end, suppose there is a 
$v\in\ker\varrho_2$ such that $v\not\in\ker\varrho_1$.  Then 
$\rho_2(vv^*)=\langle v,\varrho_2v\rangle=0$ while $\rho_1(vv^*)=
\langle v,\varrho_1 v\rangle=\|(\varrho_1)^{1/2}v\|^2>0$,
contradicting $\rho_1\ll\rho_2$ by proposition \ref{prop:abscont}.
\end{proof}

Note that by proposition \ref{prop:ac}, this lemma makes the absolute
continuity condition on the observation laws easy to verify explicitly in
the finite-dimensional setting.  In particular, the condition always holds
if $\varrho_2$ has full rank.  This is very convenient in practice: it
means that if the model is observable, we can always obtain the correct
filtered estimates as $t\to\infty$ even when the true initial state of the
system is completely unknown by choosing an initial state for the filter
of full rank.

\subsection{Homodyne detection}

For homodyne detection, we consider the observations
$$
	Y_t = U_t^*\{\sqrt{\eta}\,(A_1(t)+A_1^\dag(t))+
	\sqrt{1-\eta}\,(A_q(t)+A_q^\dag(t))\}U_t,\qquad
	\eta\in\mbox{}]0,1],~
	q_0<q;
$$
here $\eta$ is the detection efficiency, and the $q$th quadrature plays 
the role of an independent corrupting noise (we allow $q_0=q$ if $\eta=1$).
The operators $Y_t$ are self-adjoint\footnote{
	The field quadrature $A_i(t)+A_i^\dag(t)$ should be interpreted
	as the Stone generator of the appropriate Weyl operator\cite{Par92}.
  	This defines the correct domain for these
	operators on which they are self-adjoint.
} and affiliated to $\mathfrak{M}_{t]}$.  Before we can proceed, we must 
verify that the nondemolition and self-nondemolition properties hold, as 
well as the Feller property of section \ref{sec:general}.

\begin{lemma}
\label{lem:homo1}
Denote by $\mathfrak{Z}_{t]}$ the Von Neumann algebra generated by
$$
	\{Z_s:=\sqrt{\eta}\,(A_1(s)+A_1^\dag(s))+\sqrt{1-\eta}\,
	(A_q(s)+A_q^\dag(s)):s\le t\}.
$$
Then $U_T^*\mathfrak{Z}_{t]}U_T=U_t^*\mathfrak{Z}_{t]}U_t=\mathfrak{Y}_{t]}$ 
for every $0\le t\le T$.
\end{lemma}

\begin{proof}
Denote by $U_{s,t}$ ($s\le t$) the solution of the Hudson-Parthasarathy 
equation for $U_t$ with the initial condition $U_s=I$.  Then it is not 
difficult to verify that $U_{s,t}U_{r,s}=U_{r,t}$ for $r\le s\le t$, and 
that $U_{s,t}$ acts as the identity on $\mathsf{\Gamma}_{s]}$ (Thm.\ 
2.3 in Ref.\ \refcite{Bar03}).  Thus $U_{s,t}\in(\mathfrak{Z}_{s]})'$, so 
that $U_T^*\mathfrak{Z}_{t]}U_T=U_t^*U_{t,T}^*\mathfrak{Z}_{t]}U_{t,T}U_t=
U_t^*\mathfrak{Z}_{t]}U_t$.  Finally, note that any spectral projection 
$P$ of $Y_s$ (with $s\le t$) can be written as $U_s^*QU_s$ where $Q$ is a 
spectral projection of $Z_s$, so that $P=U_t^*QU_t$ also.  But the set of 
all such $Q$ generate $\mathfrak{Z}_{t]}$ and the set of all such $P$ 
generate $\mathfrak{Y}_{t]}$, so 
$U_t^*\mathfrak{Z}_{t]}U_t=\mathfrak{Y}_{t]}$.
\end{proof}

\begin{corollary}
\label{cor:homo2}
The self-nondemolition and nondemolition properties hold:
$$
	\mathfrak{Y}_{t]}\mbox{ is commutative},\qquad
	j_t(X) \in \mathfrak{Y}_{t]}'\qquad
	\forall\,t\ge 0,~X\in\mathcal{A}. 
$$
\end{corollary}

\begin{proof}
As $\mathfrak{Z}_{t]}$ is a commutative algebra and
$\mathfrak{Y}_{t]}=U_t^*\mathfrak{Z}_{t]}U_t$, evidently
$\mathfrak{Y}_{t]}$ is commutative also.  To prove the nondemolition 
property, fix $X\in\mathcal{A}$ and $P\in\mathfrak{Y}_{t]}$.
Then $[j_t(X),P]=[U_t^*(X\otimes I)U_t,U_t^*QU_t]=U_t^*[X\otimes 
I,Q]U_t=0$, as $[\mathfrak{M}_{0]},\mathfrak{Z}_{t]}]=0$.
\end{proof}

\begin{remark}
\label{rem:homofilt}
By virtue of the nondemolition and self-nondemolition properties, the 
filtering problem is well-posed.  In this setting, one can compute the 
filter explicitly as the solution of the following stochastic differential 
equation:
$$
	d\pi_t^\rho(X) = \pi_t^\rho(\mathscr{L}[X])\,dt
	+ \sqrt{\eta}\,\{
		\pi_t^\rho(L_1^*X+XL_1)	-\pi_t^\rho(L_1+L_1^*)\,
		\pi_t^\rho(X)
	\}\,d\overline{W}_t^\rho,
$$
where $d\overline{W}_t^\rho=dY_t - \sqrt{\eta}\,\pi_t^\rho(L_1+L_1^*)\,dt$
and $\pi_0^\rho(X)=\rho(X)$, see, e.g., sec.\ 5.2.4 in 
Ref.\ \refcite{Han07b}.  However, we do not need this representation of 
the filter in this paper.
\end{remark}

We must still demonstrate the remaining requirement of section
\ref{sec:general}.

\begin{lemma}
\label{lem:homo3}
For any $t_1,\ldots,t_k>0$ and bounded continuous
$f_1,\ldots,f_k:\mathbb{R}\to\mathbb{R}$,
$$
	\Phi_\rho(f_1(Y_{t_1})\cdots 
		f_k(Y_{t_k})|\mathfrak{M}_{0]}) =
	j_0(Z(t_1,\ldots,t_k,f_1,\ldots,f_k))
$$
for some $Z(t_1,\ldots,t_k,f_1,\ldots,f_k)\in\mathcal{A}$ independent of
$\rho$, and moreover
$$
	\Phi_\rho(f_1(Y_{s+t_1}-Y_s)\cdots 
		f_k(Y_{s+t_k}-Y_s)|\mathfrak{M}_{s]}) =
	j_s(Z(t_1,\ldots,t_k,f_1,\ldots,f_k))
$$
for every $s\ge 0$.
\end{lemma}

\begin{proof}
The first assertion is trivial in the current setting, as the conditional 
expectation $\Phi_\rho(~\cdot~|\mathfrak{M}_{0]})$ does not depend on 
$\rho$ and any element of $\mathfrak{M}_{0]}$ can be written as
$j_0(X)=X\otimes I$ for some $X\in\mathcal{A}$.  To prove the second 
assertion, note that
$$
	f_1(Y_{s+t_1}-Y_s)\cdots f_k(Y_{s+t_k}-Y_s) = 
	U_T^*f_1(Z_{s+t_1}-Z_s)\cdots f_k(Z_{s+t_k}-Z_s)U_T,
$$
where $T$ is chosen to be greater than $\max\{s+t_\ell:\ell=1,\ldots,k\}$.
But as $U_s,U_s^*\in\mathfrak{M}_{s]}$, we find by the module property of
the conditional expectation
\begin{multline*}
	\Phi_\rho(f_1(Y_{s+t_1}-Y_s)\cdots 
		f_k(Y_{s+t_k}-Y_s)|\mathfrak{M}_{s]}) \\
	= U_s^*\,\Phi_\rho(
		U_{s,T}^*
		f_1(Z_{s+t_1}-Z_s)\cdots f_k(Z_{s+t_k}-Z_s)
		U_{s,T}
	|\mathfrak{M}_{s]})\,U_s.
\end{multline*}
Now note that for any pair of exponential vectors 
$e(f),e(g)\in\mathsf{\Gamma}$ and $v,w\in\mathbb{C}^p$
\begin{equation*}
\begin{split}
	&\langle v\otimes e(f),
		U_{s,T}^*
		f_1(Z_{s+t_1}-Z_s)\cdots f_k(Z_{s+t_k}-Z_s)
		U_{s,T}
	~w\otimes e(g)\rangle \\
	&\qquad = \langle v\otimes e(\theta_sf),
	U_{T-s}^*f_1(Z_{t_1})\cdots f_k(Z_{t_k})U_{T-s}~
	w\otimes e(\theta_sg)\rangle~
	\langle e(f_{s]}),e(g_{s]})\rangle \\
	&\qquad = \langle v\otimes e(\theta_sf),
	f_1(Y_{t_1})\cdots f_k(Y_{t_k})~
	w\otimes e(\theta_sg)\rangle~
	\langle e(f_{s]}),e(g_{s]})\rangle,
\end{split}
\end{equation*}
where $\theta_sf(t)=f(s+t)$ and $f_{s]}$ is the restriction of $f$ to
$[0,s]$.  Hence
\begin{equation*}
\begin{split}
	&\langle v\otimes e(f),\Phi_\rho(
		U_{s,T}^*
		f_1(Z_{s+t_1}-Z_s)\cdots f_k(Z_{s+t_k}-Z_s)
		U_{s,T}|\mathfrak{M}_{s]})
	~w\otimes e(g)\rangle \\
	&\qquad =
	\langle v\otimes e(f_{s]})\otimes\xi_{[s},
		U_{s,T}^*
		f_1(Z_{s+t_1}-Z_s)
	\\ &\qquad\qquad\qquad\qquad
		\cdots f_k(Z_{s+t_k}-Z_s)
		U_{s,T}
	~w\otimes e(g_{s]})\otimes\xi_{[s}\rangle~
	\langle e(f_{[s}),e(g_{[s})\rangle \\
	&\qquad =
	\langle v\otimes\xi,
	f_1(Y_{t_1})\cdots f_k(Y_{t_k})~
	w\otimes\xi\rangle
	~\langle e(f_{s]}),e(g_{s]})\rangle 
	~\langle e(f_{[s}),e(g_{[s})\rangle \\
	&\qquad =
	\langle v\otimes\xi,
	f_1(Y_{t_1})\cdots f_k(Y_{t_k})~
	w\otimes\xi\rangle
	~\langle e(f),e(g)\rangle \\
	&\qquad = 
	\langle v\otimes e(f),\Phi_\rho(
		f_1(Y_{t_1})\cdots f_k(Y_{t_k})
	|\mathfrak{M}_{0]})
	~w\otimes e(g)\rangle.
\end{split}
\end{equation*}
The result now follows as the exponential vectors are total in 
$\mathsf{\Gamma}$.
\end{proof}

We have now completed verifying that all the requirements of section
\ref{sec:general} are met, and thus theorem \ref{thm:main} applies.  The
remainder of this section is devoted to the following problem: can one
determine directly whether the model is observable on the basis of the
coefficients $S_{ij},L_i,H$?  We will find that this is indeed the case,
and we will give an explicit algorithm to test observability. Most
of the work consists of the computation of the characteristic function of
the finite-dimensional distributions of the observation process; we employ
for this purpose a technique used by Barchielli\cite{Bar03}.

\begin{lemma}
\label{lem:barchielli1}
For any $0=t_0\le t_1\le t_2\le\cdots\le t_k$, we define
$$
	\Upsilon_{t_1,\ldots,t_k}(\lambda_1,\ldots,\lambda_k) =
	\Phi_\rho(
		e^{\sum_{\ell=1}^k\{
			i\lambda_\ell (Y_{t_\ell}-Y_{t_\ell-1})
			+ \frac{1}{2}\lambda_\ell^2(t_\ell-t_{\ell-1})
		\}}
	|\mathfrak{M}_{0]}).
$$
Then we can write
$$
	\Upsilon_{t_1,\ldots,t_k}(\lambda_1,\ldots,\lambda_k) =
	e^{(\mathscr{L}+i\lambda_1\sqrt{\eta}\,\mathscr{K})t_1}\cdots
	e^{(\mathscr{L}+i\lambda_k\sqrt{\eta}\,\mathscr{K})(t_k-t_{k-1})}I,
$$
where $\mathscr{K}[X] = L_1^*X+XL_1$.
\end{lemma}

\begin{proof}
Let $\kappa:[0,\infty[\mbox{}\to\mathbb{R}$ be locally bounded and 
measurable and define
$$
	\Xi_t(\kappa) = 
		U_t^*\,
		\exp\left(i\int_0^t \kappa(s)\,dZ_s
		+\frac{1}{2}\int_0^t \kappa(s)^2\,ds\right)
		U_t.
$$
Using the quantum It\^o rules, we find that
\begin{multline*}
	d\Xi_t(\kappa) =
	i\kappa(t)\sqrt{\eta}\,\Xi_t(\kappa)
	\sum_{k=1}^{q_0} \{j_t(S_{1k})\,dA_k(t) +
		j_t(S_{1k}^*)\,dA_k^\dag(t)\}
	\\ + i\kappa(t)\sqrt{\eta}\,j_t(L_1+L_1^*)\,\Xi_t(\kappa)\,dt 
	+ i\kappa(t)\sqrt{1-\eta}\,\Xi_t(\kappa)\,(dA_q(t)+dA_q^\dag(t)).
\end{multline*}
Similarly, we find that
\begin{multline*}
	dj_t(X) = j_t(\mathscr{L}[X])\,dt +
	\sum_{i,j,k=1}^{q_0}(S_{ki}^*XS_{kj}-\delta_{ij}X)\,d\Lambda_{ij}(t)
	\\
	+ \sum_{i,k=1}^{q_0}\left\{
		j_t(S_{ki}^*[X,L_k])\,dA_i^\dag(t) +
		j_t([L_k^*,X]S_{ki})\,dA_i(t)
	\right\}.
\end{multline*}
Using the quantum It\^o rules once more and retaining only the time 
integrals,
$$
	j_t(X)\Xi_t(\kappa) = X + \int_0^t\left\{
		j_s(\mathscr{L}[X]) +
		i\kappa(s)\sqrt{\eta}\,j_s(\mathscr{K}[X])
	\right\}
	\Xi_s(\kappa)\,ds + \mbox{martingales}.
$$
Thus evidently, if we define $\Upsilon_t(\kappa,X)=\Phi_\rho(
j_t(X)\Xi_t(\kappa)|\mathfrak{M}_{0]})$, then
$$
	\frac{d}{dt}\,\Upsilon_t(\kappa,X) =
	\Upsilon_t(\kappa,\mathscr{L}[X] + i\kappa(t)
		\sqrt{\eta}\,\mathscr{K}[X]).
$$
The result now follows directly by setting 
$$
	\kappa(s) = 
	\lambda_1I_{[0,t_1[}(s)+
	\lambda_2I_{[t_1,t_2[}(s)+\cdots+
	\lambda_kI_{[t_{k-1},t_k[}(s),
$$
then solving the equation for $\Upsilon_t(\kappa,X)$ with $X=I$.
\end{proof}

\begin{proposition}
\label{prop:homoobs}
The observable space $\mathcal{O}$ can be characterized as
$$
	\mathcal{O}=\mathrm{span}\{
		\mathscr{L}^{c_1}\mathscr{K}^{d_1}\mathscr{L}^{c_2}
		\cdots
		\mathscr{L}^{c_k}\mathscr{K}^{d_k}I:
	k,c_i,d_i\ge 0\}.
$$
In particular, $\mathcal{O}$ is the smallest linear subspace of 
$\mathcal{A}$ that contains $I$ and is invariant under the action of
$\mathscr{L}$ and $\mathscr{K}$.  The model is observable if and 
only if $\dim\mathcal{O}=p^2$.
\end{proposition}

\begin{proof}
First, we claim that $\mathcal{O}$ coincides with the linear span of 
$\Upsilon_{t_1,\ldots,t_k}(\lambda_1,\ldots,\lambda_k)$ for all
$t_1,\ldots,t_k,\lambda_1,\ldots,\lambda_k$.  To see this, note that
the characteristic function of the joint distribution of
$Y_{t_1},\ldots,Y_{t_k}$ under the state $\Phi_\rho$ is precisely
$\rho(\Upsilon_{t_1,\ldots,t_k}(\lambda_1,\ldots,\lambda_k))$ (up to
a constant factor).  As the finite dimensional distributions determine
the law of the observations, we have $\rho_1\backsim\rho_2$ if and 
only if
$$
	\rho_1(\Upsilon_{t_1,\ldots,t_k}(\lambda_1,\ldots,\lambda_k))
	=
	\rho_2(\Upsilon_{t_1,\ldots,t_k}(\lambda_1,\ldots,\lambda_k))
	\qquad\forall\,t_1,\ldots,t_k,\lambda_1,\ldots,\lambda_k.
$$
Thus evidently every element of the linear span of
$\Upsilon_{t_1,\ldots,t_k}(\lambda_1,\ldots, \lambda_k)$ is in
$\mathcal{O}$. Conversely, suppose that $X\in\mathcal{O}$ is not in the 
linear span of $\Upsilon_{t_1,\ldots,t_k}(\lambda_1,\ldots,\lambda_k)$; 
then there must exist an $Y\in M_p$ such that $\mathrm{Tr}[Y^*X]\ne 0$, 
but $\mathrm{Tr}[Y^*Z]=0$ for all $Z$ in the linear span of
$\Upsilon_{t_1,\ldots,t_k}(\lambda_1,\ldots, \lambda_k)$.  But writing $Y$ 
as $\alpha(\varrho_a-\varrho_b)+i\beta(\varrho_c-\varrho_d)$ with
$\alpha,\beta\in\mathbb{R}$ and $\varrho_a,\ldots,\varrho_d$ density
matrices corresponding to states $\rho_a,\ldots,\rho_d$, we find that 
either $\rho_a(X)\ne\rho_b(X)$, or $\rho_c(X)\ne\rho_d(X)$, while 
nonetheless $\rho_a\backsim\rho_b$ and $\rho_c\backsim\rho_d$.  We thus 
have a contradiction, and the claim is established. 

We now claim that the linear span of 
$\Upsilon_{t_1,\ldots,t_k}(\lambda_1,\ldots, \lambda_k)$ coincides with 
the linear span of $\mathscr{L}^{c_1}\mathscr{K}^{d_1}
\cdots\mathscr{L}^{c_k}\mathscr{K}^{d_k}I$.  First, note that any element 
of the latter form can be obtained from 
$\Upsilon_{t_1,\ldots,t_k}(\lambda_1,\ldots, \lambda_k)$ by taking 
derivatives with respect to $t_i$ and $\lambda_i$.  This means, in 
particular, that any element of the latter form is in the closure of the 
linear span of $\Upsilon_{t_1,\ldots,t_k}(\lambda_1,\ldots, \lambda_k)$.
But we are working in finite dimensions, so the linear span is already 
closed.  It remains to show that any $\Upsilon_{t_1,\ldots,t_k}(\lambda_1,
\ldots, \lambda_k)$ is in the linear span of elements of the form
$\mathscr{L}^{c_1}\mathscr{K}^{d_1}\cdots\mathscr{L}^{c_k}\mathscr{K}^{d_k}I$.
This is an immediate consequence of the Cayley-Hamilton theorem, and the 
claim is established.

Finally, we must show that $\mathcal{O}$ is the smallest linear subspace 
of $\mathcal{A}$ that contains $I$ and is invariant under the action of
$\mathscr{L}$ and $\mathscr{K}$.  Note that $\mathcal{O}$ is clearly 
invariant under $\mathscr{L}$ and $\mathscr{K}$ and contains $I$, so the 
smallest linear subspace such that this holds is contained in 
$\mathcal{O}$.  Conversely, any element of $\mathcal{O}$ can be generated 
by applying $\mathscr{L}$ and $\mathscr{K}$ to $I$ finitely many times
and taking finitely many linear combinations, and the smallest subspace
must contain at least these elements.  This establishes the claim.  Note 
that the model is observable if and only if $\mathcal{O}=\mathcal{A}$, 
which is clearly equivalent to $\dim\mathcal{O}=p^2$.  The proof is 
complete.
\end{proof}

Using this characterization of $\mathcal{O}$ we can construct and explicit 
algorithm for verifying observability.  To this end, we define the linear 
spaces $\mathcal{Z}_n\subset\mathcal{A}$ by
$$
	\mathcal{Z}_0=\mathrm{span}\{I\},\qquad
	\mathcal{Z}_n=\mathrm{span}\{\mathcal{Z}_{n-1},~
		\mathscr{L}[\mathcal{Z}_{n-1}],~
		\mathscr{K}[\mathcal{Z}_{n-1}]\},\quad n\ge 1.
$$
Clearly every element of $\mathcal{O}$ will be in $\mathcal{Z}_n$ for some
finite $n$.  Moreover, if $\mathcal{Z}_n=\mathcal{Z}_{n+1}$ for some
$n=m$, then it is true for all $n>m$, and in particular
$\mathcal{Z}_m=\mathcal{O}$.  But this will always be the case for some
finite $n$: after all, the linear spaces $\mathcal{Z}_n$ grow with $n$,
but $\dim\mathcal{Z}_n$ can not exceed $p^2$.  Hence this construction is
guaranteed to yield $\mathcal{O}$ in a finite number of steps.  To
implement the procedure as a computational algorithm, one could start with 
$\{I\}$ in the first step, then apply the Gram-Schmidt procedure at every 
iteration $n$ to obtain a basis for $\mathcal{Z}_n$.

\subsection{Photon counting}

In the photon counting case, we consider the observations
$$
	Y_t = U_t^*\{
		\eta\,\Lambda_{11}(t) + (1-\eta)\,\Lambda_{qq}(t)
		+\sqrt{\eta(1-\eta)}\,(\Lambda_{1q}(t)+\Lambda_{q1}(t))
	\}U_t,
$$
where $\eta\in\mbox{}]0,1]$ is again the detection efficiency and $q_0<q$.  
Once again $Y_t$ is self-adjoint and affiliated to $\mathfrak{M}_{t]}$, 
and we must verify the various properties of section \ref{sec:general}.  
The proofs of these properties are identical, however, to the homodyne 
case, so there is no need to repeat them.  We only collect here the 
required facts.

\begin{lemma}
Denote by $\mathfrak{N}_{t]}$ the Von Neumann algebra generated by
$$
	\{N_s:=
		\eta\,\Lambda_{11}(s) + (1-\eta)\,\Lambda_{qq}(s)
		+\sqrt{\eta(1-\eta)}\,(\Lambda_{1q}(s)+\Lambda_{q1}(s))
	:s\le t\}.
$$
Then $U_T^*\mathfrak{N}_{t]}U_T=U_t^*\mathfrak{N}_{t]}U_t=\mathfrak{Y}_{t]}$ 
for every $0\le t\le T$.  In particular, the self-nondemolition and 
nondemolition properties hold:
$$
	\mathfrak{Y}_{t]}\mbox{ is commutative},\qquad
	j_t(X) \in \mathfrak{Y}_{t]}'\qquad
	\forall\,t\ge 0,~X\in\mathcal{A}. 
$$
Moreover, for any $t_1,\ldots,t_k>0$ and bounded continuous
$f_1,\ldots,f_k:\mathbb{R}\to\mathbb{R}$,
$$
	\Phi_\rho(f_1(Y_{t_1})\cdots 
		f_k(Y_{t_k})|\mathfrak{M}_{0]}) =
	j_0(Z(t_1,\ldots,t_k,f_1,\ldots,f_k))
$$
for some $Z(t_1,\ldots,t_k,f_1,\ldots,f_k)\in\mathcal{A}$ independent of
$\rho$, and moreover
$$
	\Phi_\rho(f_1(Y_{s+t_1}-Y_s)\cdots 
		f_k(Y_{s+t_k}-Y_s)|\mathfrak{M}_{s]}) =
	j_s(Z(t_1,\ldots,t_k,f_1,\ldots,f_k))
$$
for every $s\ge 0$.
\end{lemma}

\begin{proof}
The proofs of these facts are identical to the proofs of lemma
\ref{lem:homo1}, corollary \ref{cor:homo2}, and lemma \ref{lem:homo3},
and are thus omitted here.
\end{proof}

\begin{remark}
Also in this setting one can compute the filter explicitly as the solution 
of a stochastic differential equation driven by the observations:
$$
	d\pi_t^\rho(X) = \pi_t^\rho(\mathscr{L}[X])\,dt
	+ \left[
		\frac{\pi_t^\rho(L_1^*XL_1)}{\pi_t^\rho(L_1^*L_1)}-
		\pi_t^\rho(X)
	\right](dY_t - \eta\,\pi_t^\rho(L_1^*L_1)\,dt),
$$
where $\pi_0^\rho(X)=\rho(X)$. We will not need this representation of 
the filter in this paper.
\end{remark}

To proceed, we must adapt lemma \ref{lem:barchielli1} to the current 
setting.

\begin{lemma}
For any $0=t_0\le t_1\le t_2\le\cdots\le t_k$, we define
$$
	\Upsilon_{t_1,\ldots,t_k}(\lambda_1,\ldots,\lambda_k) =
	\Phi_\rho(
		e^{\sum_{\ell=1}^k\{
			i\lambda_\ell (Y_{t_\ell}-Y_{t_\ell-1})
		\}}
	|\mathfrak{M}_{0]}).
$$
Then we can write
$$
	\Upsilon_{t_1,\ldots,t_k}(\lambda_1,\ldots,\lambda_k) =
	e^{(\mathscr{L}+(e^{i\lambda_1}-1)\eta\mathscr{J})t_1}\cdots
	e^{(\mathscr{L}+(e^{i\lambda_k}-1)\eta\mathscr{J})
		(t_k-t_{k-1})}I,
$$
where $\mathscr{J}[X] = L_1^*XL_1$.
\end{lemma}

\begin{proof}
Let $\kappa:[0,\infty[\mbox{}\to\mathbb{R}$ be locally bounded and 
measurable and define
$$
	\Xi_t(\kappa) = 
		U_t^*\,
		\exp\left(i\int_0^t \kappa(s)\,dN_s\right)
		U_t.
$$
Using the quantum It\^o rules, we find that
\begin{equation*}
\begin{split}
	d\Xi_t(\kappa) &=
	\eta\,(e^{i\kappa(t)}-1)\Xi_t(\kappa)
	\sum_{i,j=1}^{q_0} j_t(S_{1i}^*S_{1j})\,d\Lambda_{ij}(t) \\
	&\quad
	+ (1-\eta)\,(e^{i\kappa(t)}-1)\Xi_t(\kappa)\,d\Lambda_{qq}(t) \\
	&\quad
	+ \sqrt{\eta(1-\eta)}\,(e^{i\kappa(t)}-1)\Xi_t(\kappa)
	\sum_{i=1}^{q_0}\{
		j_t(S_{1i}^*)\,d\Lambda_{iq}(t)
		+j_t(S_{1i})\,d\Lambda_{qi}(t)
	\} \\
	&\quad
	+ \eta\,(e^{i\kappa(t)}-1)\Xi_t(\kappa)\sum_{i=1}^{q_0} 
		\{j_t(S_{1i}^*L_1)\,dA_i^\dag(t)
		+ j_t(L_1^*S_{1i})\,dA_i(t)\} \\
	&\quad
	+ \sqrt{\eta(1-\eta)}\,(e^{i\kappa(t)}-1)\Xi_t(\kappa)
	\sum_{i=1}^{q_0}\{
		j_t(L_1^*)\,dA_q(t)
		+ j_t(L_1)\,dA_q^\dag(t)
	\} \\
	&\quad
	+ \eta\,(e^{i\kappa(t)}-1)\Xi_t(\kappa)\,j_t(L_1^*L_1)\,dt.
\end{split}
\end{equation*}
Using the quantum It\^o rules once more and retaining only the time 
integrals,
$$
	j_t(X)\Xi_t(\kappa) = X + \int_0^t\{
	j_s(\mathscr{L}[X])
	+ (e^{i\kappa(s)}-1)\eta\,j_s(\mathscr{J}[X])
	\}\,\Xi_s(\kappa)\,ds
	+ \mbox{martingales}.
$$
Thus evidently, if we define $\Upsilon_t(\kappa,X)=\Phi_\rho(
j_t(X)\Xi_t(\kappa)|\mathfrak{M}_{0]})$, then
$$
	\frac{d}{dt}\,\Upsilon_t(\kappa,X) =
	\Upsilon_t(\kappa,
		\mathscr{L}[X] + 
		(e^{i\kappa(t)}-1)\eta\mathscr{J}[X]).
$$
The result now follows directly by setting 
$$
	\kappa(s) = 
	\lambda_1I_{[0,t_1[}(s)+
	\lambda_2I_{[t_1,t_2[}(s)+\cdots+
	\lambda_kI_{[t_{k-1},t_k[}(s),
$$
then solving the equation for $\Upsilon_t(\kappa,X)$ with $X=I$.
\end{proof}

The following result now follows precisely as before.

\begin{proposition}
The observable space $\mathcal{O}$ can be characterized as
$$
	\mathcal{O}=\mathrm{span}\{
		\mathscr{L}^{c_1}\mathscr{J}^{d_1}\mathscr{L}^{c_2}
		\cdots
		\mathscr{L}^{c_k}\mathscr{J}^{d_k}I:
	k,c_i,d_i\ge 0\}.
$$
In particular, $\mathcal{O}$ is the smallest linear subspace of 
$\mathcal{A}$ that contains $I$ and is invariant under the action of
$\mathscr{L}$ and $\mathscr{J}$.  The model is observable if and 
only if $\dim\mathcal{O}=p^2$.
\end{proposition}

\begin{proof}
The proof is identical to that of proposition \ref{prop:homoobs}.
\end{proof}

\subsection{Some remarks}

In this section, we have obtained precise characterizations of when a 
homodyne detection or photon counting model is observable (when the 
initial system is finite dimensional).  This yields a simple algorithm to 
test observability, from which stability of the filter follows directly 
due to theorem \ref{thm:main}.  Even in the absence of observability, 
however, one can say something about the stability of certain observables 
using corollary \ref{cor:maingen}.  The simplest such result is the 
following.

\begin{corollary}
For the homodyne detection model (resp.\ photon counting model), the 
observable $M=L_1+L_1^*$ (resp.\ $L_1^*L_1$) is always stable in the sense 
that
$$
	\Phi_{\rho_1}(
		|\pi^{\rho_1}_t(M)-\pi^{\rho_2}_t(M)|)
	\xrightarrow{t\to\infty}0
$$
whenever the law of the observations under $\Phi_{\rho_1}$ is
absolutely continuous with respect to the law of the observations under
$\Phi_{\rho_2}$.
\end{corollary}

\begin{proof}
This is immediate from corollary \ref{cor:maingen} and the fact that
$M=L_1+L_1^*=\mathscr{K}[I]$ (resp.\ $M=L_1^*L_1=\mathscr{J}[I]$) is 
clearly in $\mathcal{O}$.  
\end{proof}

In the physics literature, the observable $M$ in this corollary is
sometimes called the \textit{measurement observable}.  The fact that the
measurement observable is always stable regardless of any other properties
of the model was established for the homodyne detection case in \S 5.3.2 
of Ref.\ \refcite{Han07b} using a different method.

We conclude this section with an example that highlights the importance of 
the absolute continuity of the observations in our results.

\begin{example}
We consider the homodyne detection model, and let us choose $q_0=1$,
$S_{11}=I$, $H=0$, and $L_1=F/2$ with $F=\mathrm{diag}\{1,2,\ldots,p\}$.
By the previous corollary, the measurement observable $M=F$ is stable in 
the sense that
$$
	\Phi_{\rho_1}(
		|\pi^{\rho_1}_t(F)-\pi^{\rho_2}_t(F)|)
	\xrightarrow{t\to\infty}0
$$
whenever the observations are absolutely continuous as required by theorem
\ref{thm:main}.  It is easily verified, however, that any state $\rho$
with density matrix of the form
$\varrho=\mathrm{diag}\{0,\ldots,0,1,0,\ldots,0\}$ is a fixed point for
the filtering equation in remark \ref{rem:homofilt}. Hence
$\Phi_{\rho_1}(|\pi^{\rho_1}_t(F)-\pi^{\rho_2}_t(F)|) \not\to 0$ when
$\rho_1,\rho_2$ are two different states of this form. Evidently the
absolute continuity requirement is essential.  We refer to Ref.\ 
\refcite{Han07} for a discussion of the connection between the weakening 
of the absolute continuity requirement and the notion of controllability 
in the classical setting.  
\end{example}

\bibliographystyle{acm}
\bibliography{ref}

\begin{thebibliography}{10}

\bibitem{AFL82}
L.~Accardi, A.~Frigerio, and J.~T. Lewis.
\newblock Quantum stochastic processes.
\newblock {\em Publ. Res. Inst. Math. Sci.}, 18:97--133, 1982.

\bibitem{Bar03}
A.~Barchielli.
\newblock Continual measurements in quantum mechanics and quantum stochastic
  calculus.
\newblock In S.~Attal, A.~Joye, and C.-A. Pillet, editors, {\em Open Quantum
  Systems III: Recent Developments}, pages 207--292. Springer, 2006.

\bibitem{Bel92}
V.~P. Belavkin.
\newblock Quantum stochastic calculus and quantum nonlinear filtering.
\newblock {\em J. Multivar. Anal.}, 42:171--201, 1992.

\bibitem{BHJ07}
L.~Bouten, R.~{van Handel}, and M.~R. James.
\newblock An introduction to quantum filtering.
\newblock {\em SIAM J. Control Optim.}, 46:2199--2241, 2007.

\bibitem{kushner1}
A.~Budhiraja and H.~J. Kushner.
\newblock Robustness of nonlinear filters over the infinite time interval.
\newblock {\em SIAM J. Control Optim.}, 36:1618--1637, 1998.

\bibitem{CV07}
P.~Chigansky and R.~van Handel.
\newblock Model robustness of finite state nonlinear filtering over the
  infinite time horizon.
\newblock {\em Ann. Appl. Probab.}, 17:688--715, 2007.

\bibitem{Con85}
J.~B. Conway.
\newblock {\em A course in functional analysis}.
\newblock Springer-Verlag, 1985.

\bibitem{CR09}
D.~Crisan and B.~Rozovsky, editors.
\newblock {\em The Oxford University Handbook of Nonlinear Filtering}.
\newblock Oxford University Press, 2009.
\newblock To appear.

\bibitem{Dix69}
J.~Dixmier.
\newblock {\em Les alg\`ebres d'op\'erateurs dans l'espace hilbertien
  (alg\`ebres de von {N}eumann)}.
\newblock Gauthier-Villars \'Editeur, Paris, 1969.
\newblock Deuxi\`eme \'edition, revue et augment\'ee, Cahiers Scientifiques,
  Fasc. XXV.

\bibitem{Fag99}
F.~Fagnola.
\newblock Quantum {M}arkov semigroups and quantum flows.
\newblock {\em Proyecciones}, 18:1--144, 1999.

\bibitem{GaZ04}
C.~Gardiner and P.~Zoller.
\newblock {\em Quantum Noise}.
\newblock Springer, third edition, 2004.

\bibitem{Gud79}
S.~P. Gudder.
\newblock A {R}adon-{N}ikod\'ym theorem for {$\ast $}-algebras.
\newblock {\em Pacific J. Math.}, 80:141--149, 1979.

\bibitem{Nau74}
J.~Naudts.
\newblock A generalised entropy function.
\newblock {\em Comm. Math. Phys.}, 37:175--182, 1974.

\bibitem{Nie83}
G.~Niestegge.
\newblock Absolute continuity for linear forms on {$B\sp{\ast} $}-algebras and
  a {R}adon-{N}ikod\'ym type theorem (quadratic version).
\newblock {\em Rend. Circ. Mat. Palermo (2)}, 32:358--376, 1983.

\bibitem{OP93}
M.~Ohya and D.~Petz.
\newblock {\em Quantum entropy and its use}.
\newblock Springer-Verlag, Berlin, 1993.

\bibitem{Oks98}
B.~{\O}ksendal.
\newblock {\em Stochastic differential equations}.
\newblock Springer, fifth edition, 1998.

\bibitem{Par92}
K.~R. Parthasarathy.
\newblock {\em An introduction to quantum stochastic calculus}, volume~85 of
  {\em Monographs in Mathematics}.
\newblock Birkh\"auser Verlag, Basel, 1992.

\bibitem{Sak98}
S.~Sakai.
\newblock {\em {$C\sp *$}-algebras and {$W\sp *$}-algebras}.
\newblock Classics in Mathematics. Springer-Verlag, Berlin, 1998.

\bibitem{Tak02}
M.~Takesaki.
\newblock {\em Theory of operator algebras. {I}}, volume 124 of {\em
  Encyclopaedia of Mathematical Sciences}.
\newblock Springer-Verlag, Berlin, 2002.

\bibitem{Han07b}
R.~{van Handel}.
\newblock Filtering, stability, and robustness, 2007.
\newblock Ph.D.\ thesis, California Institute of Technology.

\bibitem{Han07}
R.~{van Handel}.
\newblock Observability and nonlinear filtering.
\newblock {\em Probab. Th. Rel. Fields}, 2008.
\newblock To appear.

\end{thebibliography}

\end{document}